# Simple sorting algorithm test based on CUDA

*Meng Hongyu, Guo Fangjin*

UCAS, Computer Science Institute, Beijing, China

**Abstract:** With the development of computing technology, CUDA has become a very important tool. In computer programming, sorting algorithm is widely used. There are many simple sorting algorithms such as enumeration sort, bubble sort and merge sort. In this paper, we test some simple sorting algorithm based on CUDA and draw some useful conclusions.

**Key words:** enumeration sort; bubble sort; merge sort; CUDA

## 1 Introduction

In this part, we will introduce the basic principle of the sort algorithm and give an overview of the GPU.

### 1.1 Enumeration sort

Enumeration sort is a simple sort algorithm and it's also named rank sort. In this algorithm, we count every element to find out the numbers of the other elements which are less than this element. Therefore, we can know the order of the element in the array. To explain more clearly, We assume that the array is a[1], a[2], ... , a[n]. Then, we compare the a[1] with a[2],a[3],…,a[n], and we count the numbers of elements which are less than a[1]. For example, if we counts 'k'. After that, we write a[1] in the b[k+1]. Array b is an orderly array. The next step, we compare the a[2] with a[1],a[3]…a[n]. Keep this step again until visit all the element. At last, the array b is the array which all the elements are in the right order. The codes are as the follows:

```
for ( int i = 0; i < N; i++ ){
   int k = 0;
     for ( int j = 0; j < N; j++ )
       if ( a[i] > a[j] )
          k ++;
     b[k] = a[i];
}
```

### 1.2 Bubble sort

Bubble sort is a sort algorithm based on compare. We repeatedly visit to array and compare two elements at one time. If the two elements are in wrong order, then we exchange them. It is very important for us to visit the array until there is no longer need to exchange. We still assume that the array is a[1], a[2], ... , a[n]. At first, we compare a[1] with a[2]. If a[2] is less than the a[1], that is a wrong order. We exchange a[1] and a[2]. Then, we keep comparing a[2] with a[3], and if a[3] is less than a[2], we exchange them. Keep this steps continue. When we have compared all the elements, we start compare a[1] with a[2] over again, and then we compare a[2] and a[3]. But in this step, we don't need to compare a[n-1] and a[n]. That means we just compare until a[n-1]. In the follows, we compare the two elements until a[n-2] and so on. At last, the a[i] array is the right array we want. The codes are as the follows:

```
for (i=1;i<N;i++)
   for(j=0;j<N-i;j++)
     if(a[j]>a[j+1])  {
           temp=a[j];
           a[j]=a[j+1];
           a[j+1]=temp;
             }
```

### 1.3 Merge sort

Merge sort is a sort algorithm based on divide and conquer. First, we divided the large array into two small arrays. So, if you want to a big orderly array, you must get two small arrays. Then, we can continue divided the small arrays. This process, we can use recursion to express. We just need to order the very small array. At last, the very important steps, we should merge all the small arrays into big arrays. Finally, we can get an orderly array, that's just we want. The codes are as the follows:

```
void Merge(int arr[],int low,int mid,int high) {
  int i=low,j=mid+1,k=0;
  int *temp=new(nothrow) int[high-low+1];
    while(i<=mid&&j<=high)  {
         if(arr[i]<=arr[j])
             temp[k++]=arr[i++];
         else
             temp[k++]=arr[j++];
       }
```

```
        while(i<=mid)
            temp[k++]=arr[i++];
        while(j<=high)
            temp[k++]=arr[j++];
        for(i=low,k=0;i<=high;i++,k++)
            arr[i]=temp[k];
        delete [] temp;
}
void MergeSort(int arr[],int n) {
    int size=1,low,mid,high;
    while(size<=n-1) {
        low=0;
        while(low+size<=n-1){
            mid=low+size-1;
            high=mid+size;
            if(high>n-1)

            high=n-1;
            Merge(arr,low,mid,high);
            low=high+1;
        }
        size*=2;
    }
}
```

### 1.4 About GPU

GPU (Graphic Processing Unit) is a specialized electronic circuit designed to rapidly manipulate and alter memory to accelerate the creation of images in a frame buffer intended for output to a display. Though the GPU is often used for graphics processing, the power of GPU for processing is also very strong compared with CPU. However, in most cases, the GPU is just used for displaying especially in personal computers although the GPU has a stronger power in processing the graphic data. In the recently years, the NVIDIA has launched many new products which have a stronger power in floating point calculations than the basic CPU, at the same time, the programming for GPU is becoming more and more easily. The CUDA platform is designed to work in GPU. CUDA gives direct access to the GPU's virtual instruction set and parallel computational elements.

## 2 Parallel programming based on CUDA

In this section, we programmed three algorithms (enumeration sort, bubble sort and merge sort) based on CUDA platform. Here are the steps.

In the first step, we use the function of "genVector()" to generate the array randomly as follows:
```
srand(0);
for(int i=0; i<N; i++)
    x[i]=i;
for ( int i = 0; i < N; ++i ){
    int r = rand() % N, tmp;
    tmp = x[i], x[i] = x[r], x[r] = tmp;
}
```

Then, we use "cudaMalloc()" for allocating the memory in video memory and use "cudaMemcpy()" for transferring the data from host to device, which means transferring the data from our computer's memory to the video memory. After that, we call the kernel function like this: "enumerationSort<<<N/threadPerBlock,threadPerBlock>>>(a, b)", in which, "N/threadPerBlock" denotes that how much blocks we use in kernel; "threadPerBlock" denotes that how threads we use in one block; 'a' and 'b' denote the data we use in the kernel, and they are in the form of array. Besides, we use "cudaThreadSynchroize()" to make sure that all the threads have completed. We use "cudaMemcpy()" once again to transfer the data from the video memory to computer memory. After all of that, we use "cudafree()" to release the video memory.

Finally, in the kernel function, we use "cnt" as the counting number, "tid" as the index of the thread in the block, and "ttid" as the index of threads in the whole blocks. We use "val" as the present variable for comparing with the other numbers in one thread. In order to reduce the time be used for loading to the data in global memory, we use shared memory to store the data. The shared memories are independent to each other in different blocks and we define one cache, which has 512 numbers in each shared memory. After that, we begin to transfer the data from global memory to shared memory. However, in this step we only transfer 512 numbers into each cache. Then we compare the "val" with the numbers in cache, and calculate out how many numbers is less than the "val" and transfer into "cnt". In the followings, we transfer the next 512 numbers into each cache and compare them. We do this until all the numbers has been compared. At last, we assign

values to array 'b' according to the "cnt". The codes are as follows:
```
int cnt = 0;
int tid = threadIdx.x;
int ttid = blockIdx.x * threadPerBlock + tid;
int val = a[ttid];
__shared__ int cache[threadPerBlock];
for ( int i = tid; i < N; i += threadPerBlock ){
    cache[tid] = a[i];
    __syncthreads();
    for ( int j = 0; j < threadPerBlock; ++j )
      if ( val > cache[j] )
        cnt++;
    __syncthreads();
}
b[cnt] = val;
```
In order to ensure results are correct, we use function to check out. If the result is wrong, it will display "sort error". Otherwise, it will display "sort right". The codes are as follows:
```
for ( int i = 0, flag =0; i < N; ++i ){
  if ( h_b[i] != i ){
    cout << "sort error" << i << "\n";
    flag = 1;
    break;
    }
}
if (flag==0)
    cout << "sort right\n";
```
To compare the results between GPU and CPU, We use some functions such as "cudaEvent_t","cudaEventSynchronize","cuda EventElapsedTime",'cudaEventElapsedTime' and "cudaEventD-estroy" to calculate the time.

## 3 Results

We run the program in our computer and get some results. The computer configuration environment is as follows:
CPU: Inter® Core™ i3-2310M CPU @ 2.10 GHZ
System: Microsoft Windows 7
RAM: 6.00 GB
GPU: GeForce GT 520M
CUDA: CUDA v-7.0
We test several times to compare CPU and GPU's results in different conditions.

```
gpu:sort right
cpu-rank:sort right
cpu-bubble:sort right
cpu-merge:sort right
host-rank:1.02179
host-bubble:0.593408
host-merge:0.094496
device: 0.048096
accelerated : 0.973696
Acc ratio: 21.2448
```

Fig. 1: sort 512 random numbers, the results between CPU and GPU

Symbol Description:
Accelerated: device sort time sub host sort time. The acc ratio is divided the time of rank sort cost on CPU by the time of GPU cost. In our program, the complexity of rank sort is as same as the codes on GPU, so we just compare the two results to conclude our speedup.

With the numbers of data increase, what happened of the results?

```
gpu:sort right
cpu-rank:sort right
cpu-bubble:sort right
cpu-merge:sort right
host-rank:5.44275
host-bubble:2.248
host-merge:0.183904
device: 0.097408
accelerated : 5.34534
Acc ratio: 55.8758
```

Fig. 2: sort 1024 random numbers, the results between CPU and GPU

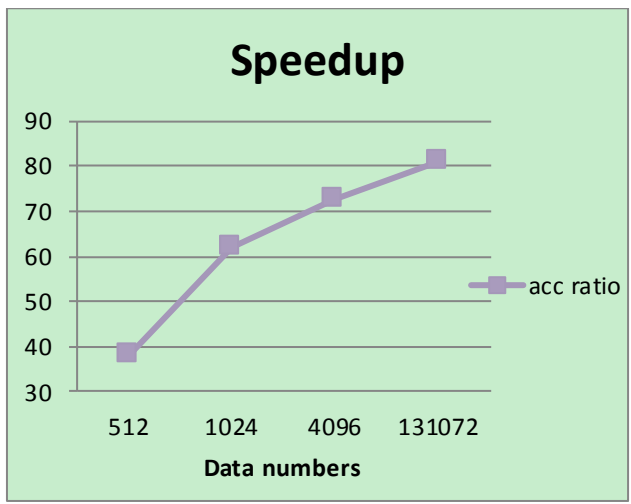

Fig. 3: sort 4096 random numbers, the results between CPU and GPU

**Speedup**

Fig. 4: The relations between data numbers and acc ratio.

From the Fig.4, we can see that with the increase of data numbers, the acc ratio is also increase. That means the speedup between CPU and GPU is more obvious. In others words, GPU are very suit for handling a large amounts of data, especial the parallel data. Compared with CPU, GPU have a great advantage in this regard.

## 4  Summary

1. The more numbers, the better result. From the above results, we can see that, when the numbers are more, the speedup is better.
2. We can reduce the time through reducing the complexity of algorithms. In rank sort, we do comparisons for n*n times, and in bubble sort, we do comparisons for n*(n-1)/2 times, and in merge sort, we just do comparisons for n*$\log_2$n times. So we use less and less time finishing the sort from rank sort to merge sort.
3. Although the GPU may cost more time than merge sort on CPU when the numbers are more than enough, we don't think that this two methods do the same work, and it makes no sense to compare them, either the bubble sort.

### References

[1] Wikipedia CUDA
http://en.wikipedia.org/wiki/Graphics_processing_unit
[2] NVIDIA CUDA SAMPLE
http://docs.nvidia.com/cuda-cuda-samples/index.html#axzz3bOt2rrsk
[3] 邓仰东，"CUDA 超大规模并行程序设计"课件，清华大学微电子学研究所